\documentclass[preprint,amsmath,amssymb,aps, prb, twocolumn, 10pt, a4paper, superscriptaddress,floatfix]{revtex4-2}

\usepackage[utf8]{inputenc}
\usepackage[english]{babel}
\usepackage[T1]{fontenc}
\usepackage{amsfonts}
\usepackage{amsthm}
\usepackage{bm}
\usepackage{csquotes}
\usepackage[normalem]{ulem}
\usepackage[x11names]{xcolor}
\PassOptionsToPackage{hyphens}{url}
\usepackage[breaklinks,colorlinks=true]{hyperref}
\usepackage[capitalise]{cleveref}
\usepackage{graphicx}
\usepackage{physics}
\usepackage{mathtools}
\usepackage{xfrac}
\usepackage{enumitem}
\usepackage{placeins}
\usepackage[caption=false]{subfig}
\usepackage{simpler-wick}
\usepackage{algorithm}
\usepackage{algpseudocode}
\usepackage{tensor}
\usepackage{newfloat}
\usepackage{cancel}
\usepackage{tikz}
\usepackage{xspace}
\usetikzlibrary{patterns.meta}
\usepackage{soul}
\usepackage{xr}

\newcommand{\nele}{n}
\newcommand{\norb}{m}
\newcommand{\ndet}{N_D}
\newcommand{\nbody}{K}
\newcommand{\normalordL}{{\{}}
\newcommand{\normalordR}{{\}}}
\newcommand{\vacuum}{{0}}
\newcommand{\methodname}{Exact Iterative Determinant--Orbital Solver\xspace}
\newcommand{\methodnameacronym}{\mbox{EIDOS}\xspace}
\newcommand{\bma}[1]{{\begin{bmatrix} #1 \end{bmatrix}}}

\newcommand{\silentcite}[1]{{\begin{@fileswfalse}\cite{#1}\end{@fileswfalse}}}
\newcommand{\externalcref}[1]{{\cref{#1}}}
\newcommand{\externaleqref}[1]{{\eqref{#1}}}

\begin{document}

\title{Precise Quantum Chemistry calculations with few Slater Determinants}

\author{Clemens Giuliani}
\email{clemens.giuliani@epfl.ch}
\affiliation{Institute of Physics, \'{E}cole Polytechnique F\'{e}d\'{e}rale de Lausanne (EPFL), CH-1015 Lausanne, Switzerland}
\author{Jannes Nys}
\email{jannys@ethz.ch}
\affiliation{Institute of Physics, \'{E}cole Polytechnique F\'{e}d\'{e}rale de Lausanne (EPFL), CH-1015 Lausanne, Switzerland}
\affiliation{Institute for Theoretical Physics, ETH Zurich, 8093 Zürich, Switzerland}
\author{Rocco Martinazzo}
\email{rocco.martinazzo@unimi.it}
\affiliation{Department of Chemistry, Universit\`a degli Studi di Milano, 20133 Milano, Italy}
\author{Giuseppe Carleo}
\email{giuseppe.carleo@epfl.ch}
\affiliation{Institute of Physics, \'{E}cole Polytechnique F\'{e}d\'{e}rale de Lausanne (EPFL), CH-1015 Lausanne, Switzerland}
\author{Riccardo Rossi}
\email{riccardo.rossi@epfl.ch}
\affiliation{Institute of Physics, \'{E}cole Polytechnique F\'{e}d\'{e}rale de Lausanne (EPFL), CH-1015 Lausanne, Switzerland}
\affiliation{CNRS, Laboratoire de Physique Th\'eorique de la Mati\`ere Condens\'ee, Sorbonne Universit\'e, 75005 Paris, France}

\begin{abstract}
Slater determinants have underpinned quantum chemistry for nearly a century, yet their full potential has remained challenging to exploit. In this work, we show that a variational wavefunction composed of a few hundred optimized non-orthogonal determinants can achieve energy accuracies comparable to the state of the art. This is obtained by introducing an optimization method that leverages the quadratic dependence of the variational energy on the orbitals of each determinant, enabling an exact iterative optimization, and uses an efficient tensor-contraction algorithm to evaluate the effective Hamiltonian with a computational cost that scales as the fourth power of the number of basis functions.  We benchmark the accuracy of the proposed method with exact full-configuration interaction results where available, and we achieve lower variational energies than coupled cluster (CCSD(T)) for several molecules in the double-zeta basis.
\end{abstract}

\maketitle

Slater Determinants (SDs) have been instrumental in shaping our understanding of quantum chemistry since their introduction nearly a century ago. In 1927, Heitler and London~\cite{Heitler1927} explained the qualitative nature of the covalent bond in $H_2$ using a wave function composed of a sum of non-orthogonal SDs built from single-atom orbitals. About two decades later, Coulson and Fischer~\cite{Coulson1949} demonstrated that the Heitler-London wave function could be significantly improved by allowing atomic orbitals to hybridize and variationally optimizing them, thereby eliminating the need to introduce ionic structures which appear unphysical for $H_2$. Subsequent developments along this line—most notably the introduction of the Generalized Valence Bond (GVB) Ansatz~\cite{Goddard1967}
 and the Spin-Coupled (SC) wave function~\cite{Gerratt1971,Cooper1991}
—marked
the foundation of modern VB theory~\cite{Lawley2007,Wu2011a}, and played a key role in our current understanding of chemical bonding. In VB theory, bonds emerge from resonating chemical structures — superpositions of basic wavefunctions — that are themselves spin-determined linear combinations of non-orthogonal SDs constructed with semi-localized orbitals. The compactness of the wave function is crucial for this interpretation and can only be achieved through simultaneous optimization of both the structures and the orbitals. However, this remains a challenging task compared to molecular orbital (MO)-based methods~\cite{
McWeeny1996},
which exploit orbital orthogonality to achieve unmatched numerical efficiency, at the expense of interpretability.
Achieving chemical accuracy requires too many contributing SDs when built with orthogonal orbitals, while
non-orthogonal SDs, potentially more interpretable, pose significant numerical challenges when high accuracy is required.

In view of the above,
it is not surprising that many techniques have been proposed to optimize SDs in a Configuration-Interaction (CI) expansion. Moreover, the Full Configuration Interaction (FCI) expansion—while exact within a chosen finite basis—scales exponentially with basis‐set size, rendering it impractical for all but the very smallest systems. Approximate methods can be generally divided in two classes:
(i) methods using one or multiple fixed-reference determinants that optimize a linear combination of excitations thereof, and (ii) multi-configuration determinantal methods that optimize the orbitals of the SDs, along with the CI coefficients.

A notable example in the first category is truncated CI, where excitations from a reference determinant are limited to a certain excitation level; for instance, only single and double excitations are included in CISD.
In such cases, the number of determinants is combinatorial in the level of excitations and CI truncation breaks size-consistency. The prohibitive scaling can be mitigated by selected-CI methods, which retain only a subset of relevant excitations. However, identifying the important determinants is generally a nontrivial task.
Quadratic Configuration Interaction~\cite{Pople1987_qcisd} and Coupled Cluster~\cite{Coester1960} 
methods such as CCSD(T) correct for the size-consistency by approximately including also higher excitations; however, unlike CI methods, they are non-variational and come at an increased computational cost.
The aforementioned methods in this class all use orthogonal Slater determinants. Non-orthogonal CI (NOCI)~\cite{Thom2009}, on the other hand, optimizes a linear combination of given, fixed, non-orthogonal Slater Determinants.
\begin{figure*}[ht]
\includegraphics[width=\textwidth]{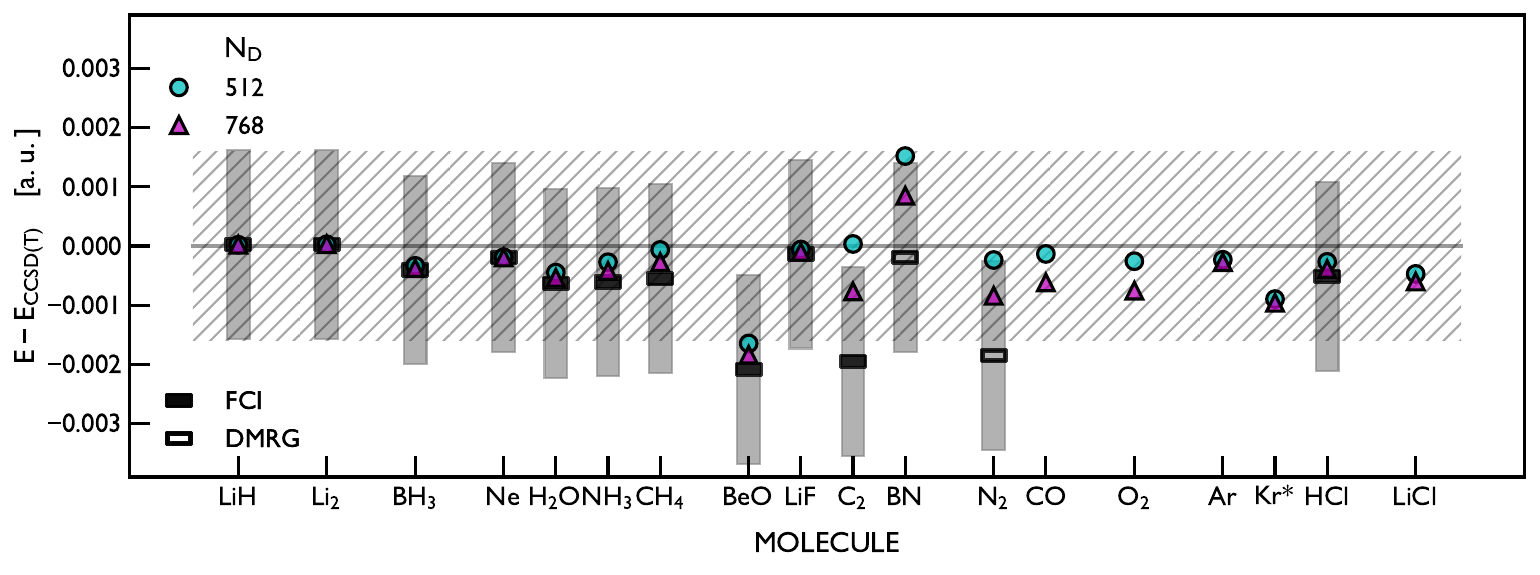}
\caption{
\textit{Energy differences with respect to CCSD(T) in the cc-pVDZ basis set.} The shaded area
\begin{tikzpicture}
\fill [black, opacity=0.3] (0,0) rectangle (0.15,0.25);
\end{tikzpicture}
indicates results within chemical accuracy (1 kcal/mol) from FCI/DMRG, and the hatched area
\begin{tikzpicture}
\fill [
       pattern={Lines[
                  distance=0.9mm,
                  angle=45,
                  line width=0.1mm
                 ]},
        pattern color=black
       ] (0,0) rectangle (0.5,0.25);
\end{tikzpicture}
from CCSD(T). Geometries and FCI reference energies (where available) are taken from Ref.~\silentcite{gao2024distributedfci}, with the exception of the N\textsubscript{2} molecule at the distance of $2.118$ $a.u.$, where we compare to DMRG results from Ref.~\silentcite{chan2004state}, and BN with our own calculation using the block2 code~\silentcite{Zhai2023_block2_dmrg}. CCSD(T) energies computed with PySCF~\silentcite{Sun2020_pyscf3}
for N\textsubscript{2} and Psi4~\silentcite{Smith2020_psi4_14} for the rest. The molecules are ordered w.r.t increasing number of electrons, grouping those with equal numbers together.
For those marked with \textsuperscript{*} we employed a particle-hole transformation.
}
\label{fig:molecules_ccpvdz}
\end{figure*}

The second class of methods--which is the most relevant in spirit to the topic of this article--consists of a variety of multi-configurational approaches that employ different strategies to optimize both the CI coefficients and the orbitals of the SDs. Thanks to the improved quality of the SDs achieved with the orbital optimization, these CI expansions are typically much more compact than those in the first class, albeit at the cost of increased computational effort.
Self-consistent field (MCSCF) methods, such as CASSCF~\cite{Siegbahn1980_casscf},
 optimize a single orbital rotation applied to all or a subset of the CI determinants.
Gradient-based approaches optimizing only the orbitals of a single determinant at a time have been proposed for two determinants~\cite{Bremond1964}, and many determinants~\cite{Koch1993_nonorthog_detopt_singled},
or Hartree-Fock-Bogoliubov wavefunctions~\cite{Schmid1989_fed,JimnezHoyos2013}.
Resonating Hartree-Fock~\cite{Fukutome1988_reshf, Ikawa1993,Tomita1996} and nonorthogonal multiconfigurational self-consistent field~\cite{Mahler2021} exploit the Thouless parametrization of the SDs~\cite{Thouless1960} to optimize all determinants and their orbitals simultaneously, using gradient-based methods. However, it can be difficult to converge~\cite{JimnezHoyos2013,Sun2024,Miller2025_reshf_svd}.
Valence Bond Self-Consistent Field (VBSCF) methods~\cite{VanLenthe1980_vbscf,
Rashid2013_vbscf} use second-order optimizers to optimize a sum of non-orthogonal determinants.
Multi-configuration time-dependent Hartree methods (MCTDH-F)~\cite{Lode2020a} extend the original MCTDH method~\cite{Meyer1990_mctdh}
to fermions and use numerical integration to apply the real or imaginary time evolution to the determinants.

In this work, we show that a sum of fewer than a thousand SDs is sufficient to accurately represent the all-electron ground-state wavefunction of small-size molecules, when the SDs are fully optimized without orthogonality constraints.
These results are achieved with a computational technique that we introduce in this article, which we call the \methodname (\methodnameacronym).  \methodnameacronym exactly optimizes the variational energy when viewed as a function of a \emph{string} of orbitals, obtained by selecting one orbital from each determinant.
\methodnameacronym\ relies on the direct parameterization of the molecular orbital coefficients
-- as opposed to the popular Thouless parameterization of the SDs~\cite{Thouless1960} --
and an efficient, analytical evaluation of the corresponding Hessian of the energy functional, resulting in a favorable $\mathcal{O}(m^4)$ scaling with the basis set size $m$ (see the Methods Section for details). We assess the accuracy of \methodnameacronym\ on several molecules in the cc-pVDZ basis, benchmarking with full-CI data where available, and with DMRG and coupled-cluster (CCSD(T))
otherwise.
We find that our variational energies are consistently below coupled cluster, and within chemical accuracy of full-CI and DMRG,
demonstrating that \methodnameacronym is competitive with state-of-the-art methods, but with significantly more favorable scaling. We additionally discuss how the number of determinant scales with bond order, and we demonstrate convergence toward the complete-basis-set limit for LiH. To test the method’s robustness, we analyze the strongly correlated regime along the N$_2$ dissociation curve, and we verify that it correctly reproduces the triplet ground state of O$_2$.

SDs also play a crucial role in many-body methods such as Quantum Monte Carlo. Compact SD expansions serve as efficient trial wavefunctions in Auxiliary-Field Quantum Monte Carlo (AFQMC)~\cite{al2006auxiliary,
lee2022twenty}
and Diffusion Monte Carlo (DMC)~\cite{scemama2016quantum}, and underpin variational ans\"atze like the Slater-Jastrow form~\cite{ceperley_monte_1977} as well as recent variants augmented with neural networks~\cite{ferminet,
hermann2020deep,liu2024neural,bortone2024simple}.

\section*{Results}
\label{sec:results}

\paragraph*{Non-orthogonal Slater-Determinant Ansatz optimized by the \methodnameacronym algorithm.}

The wave function we consider consists of a sum of unrestricted un-normalized non-orthogonal SDs\begin{equation}
    \label{eq:sum_slater_spin}
    \ket{\bar \Phi} = \sum_{I=1}^{\ndet} \hat\Phi^{(I \uparrow)}\, \hat\Phi^{(I \downarrow)} \ket{\vacuum},
\end{equation}
where $\ndet$ is the number of determinants, and
\begin{equation}
\label{eq:slater_det_def}
    \hat\Phi^{(I,\sigma)} =
    \prod_{i=1}^{\nele_\sigma}\left(  \sum_{\mu=1}^{\norb} \phi_{i, \mu}^{(I,\sigma)} \,\hat c^\dagger_{\mu,\sigma}\right),
\end{equation}
where $\phi^{(I,\sigma)}_{i}\in \mathbb C^{\norb}$ are independent ``MO orbitals'', $\nele_{\sigma}$ is the number of electrons of spin $\sigma \in \{\uparrow,\downarrow\}$, $\norb$ is the number of basis orbitals, and $\hat c_{\mu,\sigma}^\dagger$ is the creation operator of an electron in orbital $\mu$ with spin $\sigma$.

We optimize the average energy of the non-orthogonal Slater determinant wave-function ansatz of Eq.~\eqref{eq:sum_slater_spin} using the \methodnameacronym algorithm introduced in this work, as detailed in the Methods section. We briefly sketch here the main idea behind the efficiency of EIDOS, which is based on two facts: (i) the average energy of the wavefunction of~\cref{eq:sum_slater_spin} is the ratio of two quadratic functions of $\phi^{(1,\sigma)}_{1},\dots,\phi^{(\ndet,\sigma)}_{1}$ at fixed electron label $i=1$, which implies that these parameters can be optimized exactly, and the procedure repeated after an electron-label rotation in each determinant; (ii) the needed effective quadratic forms can be computed with a total computational cost of $O(\norb^4)$ with an efficient tensor-contraction strategy.

\paragraph*{Benchmark with CCSD(T), DMRG and FCI for equilibrium geometries in the cc-pVDZ basis.}
To validate the accuracy of the proposed method, we compare it against CCSD(T), and, where available, DMRG and FCI, for the all-electron ground-state energy of several molecules at equilibrium in the cc-pVDZ basis set.
In our calculations, the electron number varies from $4$ (LiH) to $20$ (for LiCl), except for the Krypton atom with $36$ electrons in $27$ orbitals, where we perform a particle-hole transformation to reduce the effective number of interacting particles to $18$.
As reported in~\cref{fig:molecules_ccpvdz}, our optimized energies from up to $\ndet=768$ Slater determinants are well within chemical accuracy (1 kcal/mol) with respect to FCI, including those at the limit of what is still tractable by current distributed FCI codes~\cite{gao2024distributedfci}.
Furthermore, our variational energies are consistently below the ones obtained with coupled cluster at CCSD(T) level, even if energy comparisons with a non-variational method could appear to not be completely fair to our method. This is significant because CCSD(T) has an asymptotic computational scaling of $O(\norb^7)$, while our method scales like $O(\norb^4)$.
For all molecules we find the expected spin symmetry for the ground state, with a maximal spin contamination at 768 determinants of $\abs*{\ev*{\hat S^2} - S(S+1)}$ $\approx 2\cdot 10^{-2}$ for BN,
and much less for the other molecules.
We also note that $768$ determinants consistently improve the results of $512$ determinants, and that the optimization steps of the EIDOS algorithm to achieve convergence are roughly independent of the number of determinants (see Supplementary Information, \externalcref{sup-additional_results}).

\begin{figure}[t]
\includegraphics[width=0.475\textwidth]{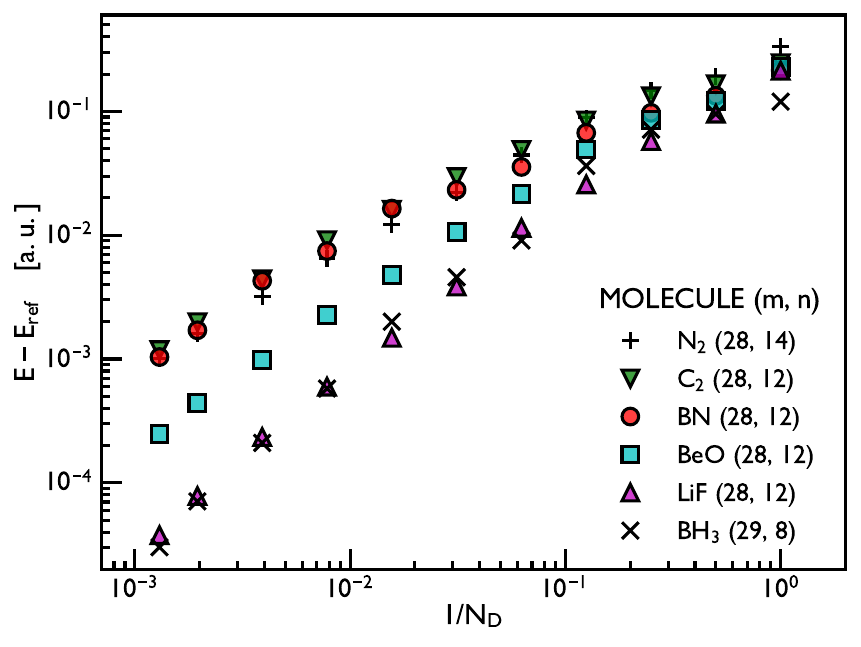}
\caption{\raggedright \textit{Scaling of the number of SDs with bond order.}
We consider here the four di-atomic molecules C\textsubscript{2}, BN, BeO and LiF with an identical number of electrons $\nele$ and basis orbitals $\norb$, and BH\textsubscript{3} and N\textsubscript{2} with similar numbers, all in the cc-pVDZ basis set.
The molecules possess different bond orders, ranging from the single bond of LiF to the triple bond of N\textsubscript{2}.
Equilibrium geometries and FCI reference energies from Ref.~\silentcite{gao2024distributedfci}, except for N\textsubscript{2} molecule at the distance of $2.118$ $a.u.$, where we compare to DMRG results from Ref. \silentcite{chan2004state}, and BN to our own DMRG calculation using the block2 code~\silentcite{Zhai2023_block2_dmrg}.
}
\label{fig:dz_bo}
\end{figure}

\paragraph*{Scaling of the number of SDs with bond order.}
In general, we expect the number of SDs needed to precisely model the ground state to increase with correlations, and, in particular, we expect a dependence on the number of valence electrons.
To verify this picture in a controlled setting, we consider the three molecules LiF, BeO and N\textsubscript{2}, which all have similar number of electrons, $\nele \in\{12,14\}$, the same number of basis orbitals in the cc-pVDZ basis set, $\norb=28$, but different bond order.
In \cref{fig:dz_bo} we report the energy error w.r.t. FCI as a function of the number of determinants, ranging from $\ndet=1$ up to $\ndet=768$, observing a smooth and consistent scaling. Moreover, the log-log scale we use in the plot graphically shows that the decay of the energy error is certainly not slower than a power-law of the inverse number of SDs.
The slope of the decay clearly decreases with the bond order/number of valence electrons, meaning that for N\textsubscript{2} we need more than an order of magnitude more determinants to attain chemical accuracy than what is needed for LiF, with BeO lying in between. The single-bonded BH\textsubscript{3} shows a similar scaling as LiF, while it is interesting to remark that the curves for C\textsubscript{2} and BN, which are considered to be doubly bonded molecules, nearly collapse on one of the triple-bonded N\textsubscript{2}.

\begin{figure}[t]
\includegraphics[width=0.475\textwidth]{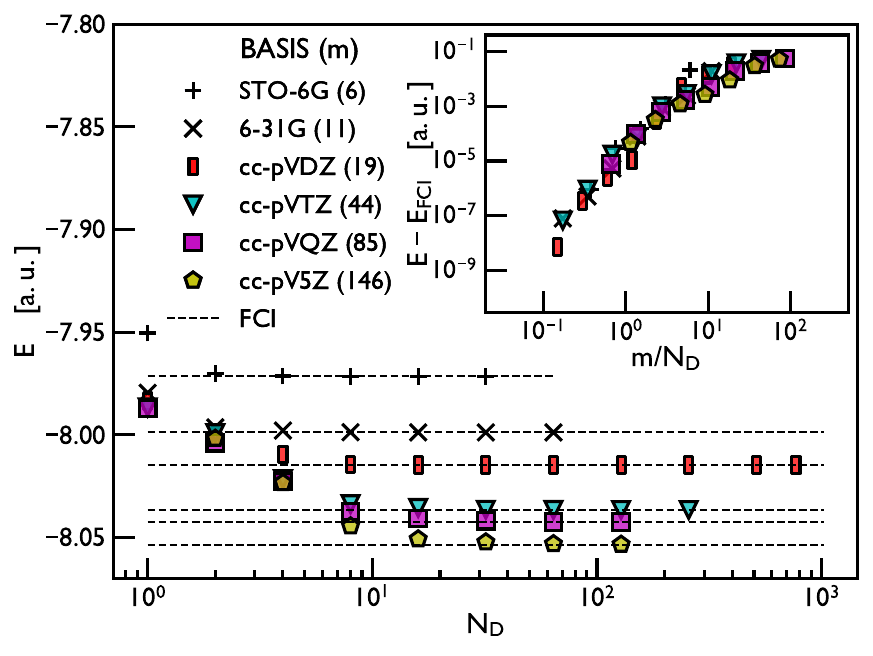}
\caption{
\textit{Convergence to the infinite-basis-set limit.} We show the absolute energy convergence as a function of the number of SDs $\ndet$ and the basis-set size $\norb$ for LiH. In the inset, we show an approximate curve collapse of the absolute error with respect to FCI in a given basis as a function of the ratio between the size of the basis and the number of SD.
Equilibrium geometry and FCI reference energies from Ref.~\silentcite{gao2024distributedfci}
(dashed lines).}
\label{fig:lih_bases}
\end{figure}

\paragraph*{Convergence to the large-basis-set limit.}
In \cref{fig:lih_bases} we consider the ground-state energy of the LiH molecule at equilibrium distance, for various basis sets of sizes ranging from $\norb=6$ to $\norb=146$ orbitals, the largest for which FCI energies are still available~\cite{gao2024distributedfci}.
The approximate curve collapse for the energy error in a given basis set as a function of the ratio of the number of SDs and the basis set size $\norb$, shown in the inset of \cref{fig:lih_bases}, suggests that the number of determinants needed to get a fixed error \textit{in a given basis} scales linearly with the basis set size. We conjecture that this might be due to the short-distance cusp conditions of the continuum-space wave function, which require having a wavefunction with an increasing number of determinants in the approach to the continuum limit.
Nevertheless, as we show in the main plot of \cref{fig:lih_bases}, the absolute energy decreases with increasing basis set size $\norb$ for a fixed number of SD $\ndet$, as does the error with respect to the true ground-state energy in the infinite-basis-set limit.

\begin{figure}[t]
\includegraphics[width=0.475\textwidth]{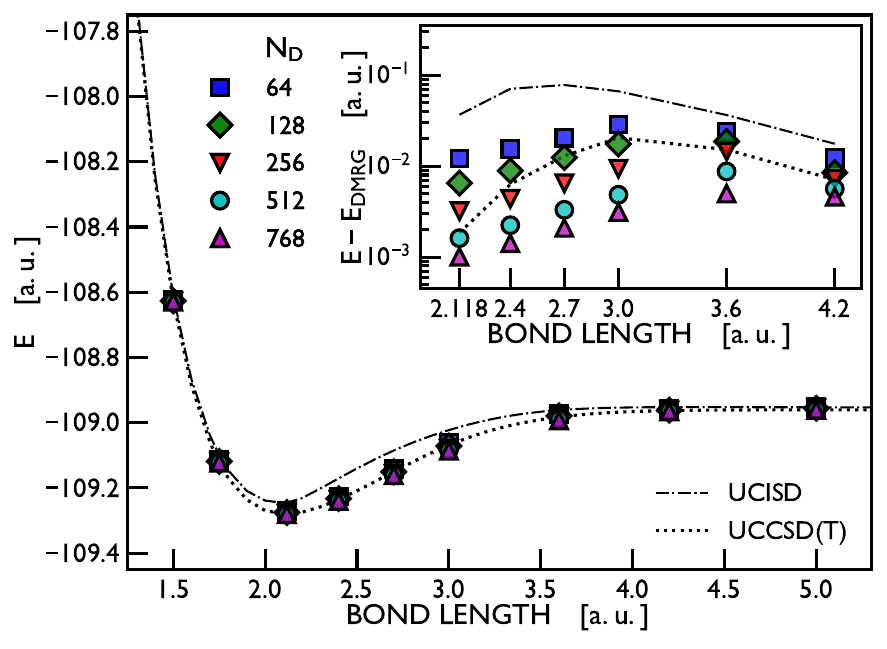}
\caption{
\textit{Dissociation curve of the N\textsubscript{2} molecule in the {cc-pVDZ} basis.} We compare our variational energy results for different numbers of determinants $\ndet$ with UCCSD(T) (dotted line) and UCISD (dash-dotted line) from our own calculations with PySCF~\silentcite{Sun2020_pyscf3}. In the inset we show the error compared to DMRG results from Ref.~\silentcite{chan2004state} for the equilibrium bond distance and larger.
}
\label{fig:n2_dissoc}
\end{figure}

\paragraph*{Capturing static correlations.}
As a prototypical example of bond breaking, we consider the Nitrogen molecule.
As its bond is stretched, the wavefunction starts to have an inherent multireference character. Single reference methods such as CISD and CCSD(T) start to struggle, converging more slowly or not converging at all~\cite{chan2004state}.
In \cref{fig:n2_dissoc} we show the dissociation curve of the N\textsubscript{2} molecule in comparison with UCISD (which uses 30724 orthogonal determinants), UCCSD(T) and DMRG.
We outperform UCISD and UCCSD(T) already with a few hundred independent Slater determinants. As can be seen in the inset, close to equilibrium we obtain energies within chemical accuracy from DMRG, however not at large distances.

\begin{figure}[t]
\includegraphics[width=0.475\textwidth]{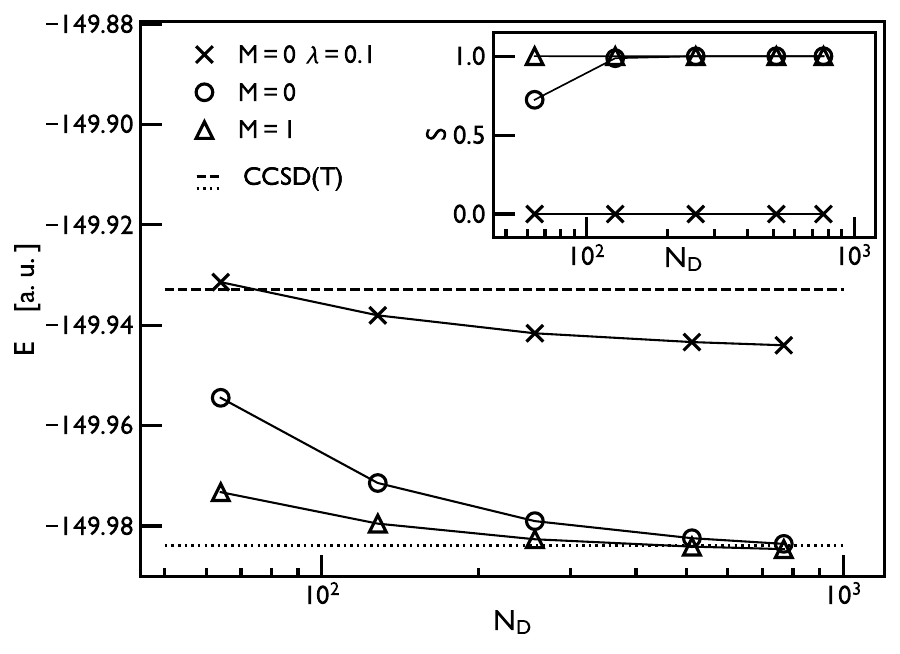}
\caption{\textit{Triplet ground state and singlet first excited state of the O\textsubscript{2} molecule in the cc-pVDZ basis set.} The singlet state is found by minimizing the energy of the Hamiltonian with an additive penalty term $\lambda\, \hat S^2$. The horizontal lines represent CCSD(T) energies for
${M=0}$, ${S=0}$ (dashed)
and ${M=1}$, ${S=1}$(dotted).
Equilibrium geometry taken from the supplementary material of Ref. \silentcite{gao2024distributedfci}. In the inset, we show the average value of the spin $S\coloneq \sqrt{\langle\hat S^2\rangle+\frac{1}{4}}-\frac{1}{2}$.}
\label{fig:o2}
\end{figure}

\paragraph*{Triplet and singlet dioxygen states.}
As an additional application, we discuss the low-energy states of molecular oxygen for different sectors of the $M$ and $S$ quantum numbers.
While ${M = \ev*{\hat S_z} = \frac{1}{2}(\nele_{\uparrow} {-} \nele_{\downarrow})}$ is directly imposed in the wavefunction of Eq.~\eqref{eq:sum_slater_spin}, in order to tune the value of $S$ in this calculation, we add a $\lambda\, \hat S^2$ penalty term to the Hamiltonian, at no extra cost during the optimization.
In \cref{fig:o2} we study the convergence of the O\textsubscript{2} to its triplet ground state, both in the $M=0$ and $M=1$ spin sectors.
For the latter we need fewer determinants to accurately represent the eigenstate, due to the smaller size of the subspace it is defined in.
Adding the $\lambda\,\hat S^2$ penalty term, with $\lambda=0.1$, we converge to the singlet state, which is known to be the first excited state of this molecule.
In the inset of \cref{fig:o2} we show that we do not observe spin contamination, apart from the lowest number of determinants in the $M=0$ sector.

\nocite{gao2024distributedfci,chan2004state,Zhai2023_block2_dmrg,Sun2020_pyscf3,Smith2020_psi4_14}
\section*{Discussion}
The main fundamental result of our work is to show that a sum of non-orthogonal Slater determinants provides a compact representation of the ground-state wave function of small-size molecules. From a computational point of view, we have numerically determined that the variational optimization of the energy of SDs achieves results comparable to state-of-the-art techniques, with a favorable scaling. 
This was obtained with a  deterministic algorithm for the variational optimization we introduce in this work, EIDOS, based on the partial-but-exact optimization of the variational ansatz when viewed as a function of one orbital per determinant, and efficient tensor contractions.  
Our results show that it is possible to achieve chemical accuracy with respect to full-CI and DMRG calculations in the correlation-consistent double-zeta basis with just a few hundred determinants, and to consistently obtain variational energies lower than coupled-cluster CCSD(T) methods, with only a $\mathcal O(\norb^4)$ scaling with the basis set size $\norb$. We have further demonstrated that the SD ansatz can be highly accurate for large basis-set sizes when compared to FCI for small molecules, and that we are able to accurately capture the strong correlations occurring at bond breaking of N\textsubscript{2}. The scaling of the energy error with the number of determinants is closely related to the bond order of the molecule, as expected from chemical intuition. In addition, we have shown that only a small number of SDs are needed to recover the expected symmetry-eigenstates of the O\textsubscript{2} molecule.

The computational bottleneck of the current version of EIDOS is the memory needed to densely store the matrices defining the quadratic forms for the average value of the energy (see Methods Section).
However, 
iterative solvers could be used instead.
Alternatively, with a similar formalism, we could optimize a subset of all the determinants at each step, or perform symmetry projection~\cite{Scuseria2011,JimnezHoyos2013}. The efficiently contractable formulae (\cref{eq:contraction_part1,eq:contraction_part2b} in the Methods section) naturally admit tensor decompositions of the Hamiltonian~\cite{Hohenstein2012_tensor_hypercontraction_1,Peng2017_tensor_decomp_for_cc_double_factorization} with the potential to further reduce the computational scaling in $\norb$.
A more physical limitation of the method is size inconsistency at fixed number of determinants: in order to have the same accuracy on a system consisting of two non-interacting parts, the number of determinants should be higher than the number of determinants used in each subsystem, as we show in \externalcref{sup-sizeconsistency}.
Another limitation is the inability of sums of Slater Determinants to represent the cusp of the wave function in the continuum limit. As discussed in the Results section, we conjecture that this is responsible for the linear increase of the number of determinants with the number of basis set orbitals, for fixed precision {\it in a given basis set}, even if the overall accuracy is increasing with larger basis set when compared to the infinite-basis-set limit. The wave-function cusp could be directly imposed by adding a Jastrow factor to the optimized Slater determinants, and further optimizing it with VMC.

A straightforward extension of this work is to evaluate the first excited states. As we have an analytical expression for wave-function overlaps, we can directly project out the approximate ground state (or any other subspace) and variationally optimize the energy, which remains a quadratic function.
It is also possible to extend the technique to time evolution, both in real and imaginary time. The approach is very similar to the one presented in this work, as the fidelity optimization is also a ratio of quadratic functions in the orbitals.
While we leave discussions of extrapolations to the infinite-determinant limit for future work, we note that we are able to compute the variance of the energy with the same computational cost as the energy, which could be used to obtain a better estimate of the ground state using the zero-variance principle.

We have introduced our approach as a standalone method.
However, it could also be used in conjunction with other established methods.
For example, it could provide a highly accurate multideterminant trial wavefunction for AFQMC and DMC calculations,
requiring a much lower number of determinants and a higher fidelity compared to the CI wavefunctions typically used.
The non-orthogonal Slater Determinant ansatz, optimized with the analytical orbital optimization strategy of the present article, typically outperforms state-of-the-art second-quantized VMC approaches using artificial neural networks.
Nevertheless, determinant-based variational Ans\"atze, augmented by neural networks, in first or second quantization, with pre-optimized determinant parts could be a promising approach for VMC.

\section*{Methods}

\paragraph*{Notation.} We use capital letters for many-body quantities:
    $\ket*{\bar\Phi}$, $\ket*{\Phi^{(I)}}$ denote $\nele$-electron wavefunctions,
    $\ndet$ is the number of determinants,
   and $I,J,K,\dots \in \{1,\dots,\ndet\}$ are determinant labels.
    For single electron quantities we use lowercase letters:
    $\nele$ is the number of electrons,
    $i,j,\dots \in \{1,\dots,\nele\}$ are electron labels,
     $\norb$ is the number of basis functions, and
    $\mu,\nu,\dots \in \{1,\dots,\norb\}$ bare basis orbitals labels.
    $\hat c^\dagger_\mu$ is the creation operator creating an electron in the $\mu$-th basis state (assumed here, for simplicity, to be orthonormal for the usual anticommutation relations to hold).

We consider the following wave-function ansatz 
\begin{equation}\label{eq-ansatz}
    \ket*{\bar \Phi} = \sum_{I=1}^{\ndet} \ket*{\Phi^{(I)}},
\end{equation}
where $\left\{\ket{\Phi^{(1)}},\ket{\Phi^{(2)}},\dots,\ket{\Phi^{(\ndet)}}\right\}$ is a set of non-orthogonal SDs.
Each Slater determinant $\ket{\Phi^{(I)}}$ is fully specified by its own sequence of ``MO orbitals'' $\left(\phi^{(I)}_{1}, \phi^{(I)}_{2},\dots, \phi^{(I)}_{\nele}\right)$, given as a column vector of coefficients, i.e. $\phi^{(I)}_i\in \mathbb C^{\norb}$.
We can build each SD $\ket{\Phi^{(I)}}$ as
\begin{equation}
    \ket*{\Phi^{(I)}}=    \left(\sum_{\mu_1=1}^m\phi_{1,\mu_1}^{(I)} \hat c^\dagger_{\mu_1}\right)
    \cdots
    \left(\sum_{\mu_{\nele}=1}^m\phi_{\nele,\mu_\nele}^{(I)} \hat c^\dagger_{\mu_\nele}\right)
    \ket{\vacuum}.
\end{equation}
In this parametrization the determinant is multilinear in the MO orbital coefficient vectors $\phi^{(I)}_i$,
and linear when viewed as a function of a single one of them.
Without loss of generality, singling out the first orbital (it is just a matter of reordering the orbitals in every determinant, and absorbing the resulting sign into one of the coefficient vectors), we have that
\begin{equation}
    \ket*{\Phi^{(I)}}
    = \sum_{\mu_1=1}^m\phi_{1,\mu_1}^{(I)} \hat c_{\mu_1}^\dagger \ket*{\Phi^{(I)}_{\text{1+}}},
\end{equation}
where $\ket*{\Phi^{(I)}_{\text{1+}}} \coloneqq \left(\sum_{\mu_2}\phi_{2,\mu_2}^{(I)} \hat c^\dagger_{\mu_2}\right)
    \cdots
    \left(\sum_{\mu_{\nele}}\phi_{\nele,\mu_\nele}^{(I)} \hat c^\dagger_{\mu_\nele}\right)
    \ket{\vacuum}$ is a ``hole''-SD, i.e., the
$I$\textsuperscript{th} SD with the first electron removed.
We can write the full wavefunction, in terms of the ``hole''-SD's as
\begin{equation}
    \ket*{\bar \Phi} = \sum_{I=1}^{\ndet} \sum_{\mu_1=1}^m\phi_{1,\mu_1}^{(I)} \hat c_{\mu_1}^\dagger \ket*{\Phi^{(I)}_{\text{1+}}}.
\end{equation}

\paragraph*{The \methodname (\methodnameacronym).} In this section, we present our method to variationally optimize the average energy of an ansatz consisting of a sum of SDs by solving at each step a linear algebra problem. We first show why it is enough to solve a linear algebra problem, and afterwards how to efficiently compute the terms appearing in the linear algebra problem. 

Our first finding  is that the energy expectation value  with the wave-function ansatz of Eq.~\eqref{eq-ansatz}, when regarded as a function of a single orbital of each SD, is simply a  ratio of two quadratic forms
\begin{equation}
\label{eq:generalized_raileigh}
    E = \frac{\mel*{\bar\Phi}{\hat H}{\bar\Phi}}{\braket*{\bar\Phi}} = \frac{v^\dagger \mathcal H v}{v^\dagger \mathcal S v},
\end{equation}
where $v \in \mathbb C^{(\ndet \times \norb)}$ , defined by
\begin{equation}
    v_{I\mu} \coloneq \phi^{(I)}_{1,\mu},
\end{equation}
is the ``vector'' (with flattened index $I\mu$, for $\qquad I\in\{1,\dots,\ndet\}$ and $\mu\in\{1,\dots,m\}$) containing the orbitals we factored out from each determinant, and
$\mathcal H, \mathcal S \in \mathbb C^{(\ndet \times \norb) \times (\ndet \times \norb)}$ are the ``effective matrices'' for respectively the Hamiltonian and the norm,  and they are given by
\begin{equation}
\label{eq:mathcalH}
\mathcal H_{I\mu,J\nu} = \mel*{\Phi^{(I)}_{\text{1+}}}{\hat c_{\mu} \hat H \hat c_{\nu}^\dagger}{\Phi^{(J)}_{\text{1+}}},
\end{equation}
\begin{equation}
\label{eq:mathcalS}
\mathcal S_{I\mu,J\nu} = \mel*{\Phi^{(I)}_{\text{1+}}}{\hat c_{\mu} \hat c_{\nu}^\dagger}{\Phi^{(J)}_{\text{1+}}},
\end{equation}
where $I,J\in\{1,\dots,\ndet\}$ and $\mu,\nu\in\{1,\dots,m\}$. 
We remark that $\mathcal H_{I\mu,J\nu}$ and $\mathcal S_{I\mu,J\nu}$ are sub-matrices of the Hessian matrices of the numerator and denominator of the energy in \cref{eq:generalized_raileigh}.
The energy \cref{eq:generalized_raileigh} takes the form of a generalized Rayleigh quotient, and can be minimized exactly w.r.t. $v$ by solving the generalized eigenvalue problem (GEV)
\begin{equation}
\label{eq:generalized_eigval}
\mathcal H v = \epsilon \,\mathcal S v
\end{equation}
taking $\mathcal O({\ndet}^3\, {\norb}^3)$ operations.
We remark that $S$ issingular, with nullspace spanned by the ``MO orbitals'' of the hole-SDs $\ket*{\Phi^{(I)}_{\text{1+}}}$. By projecting it out, the problem can be reduced from $\ndet \times \norb$ to a smaller eigenvalue problem of size $\ndet \times(\norb{-}\nele{+}1)$.

Our second finding is that the effective Hamiltonian matrices $\mathcal H$ and $\mathcal S$ can be calculated efficiently.
The hole-wavefunction matrix elements $\mel*{\Phi^{(I)}_{\text{1+}}}{\hat c_{\mu} \hat H \hat c_{\nu}^\dagger}{\Phi^{(J)}_{\text{1+}}}$ needed for $\mathcal H$ can be calculated in a numerically-exact way at an asymptotic cost of $\mathcal O({\ndet^2\, \norb}^4)$, including the case in which $\braket*{\Phi^{(I)}_{\text{1+}}}{\Phi^{(J)}_{\text{1+}}} = 0$.
Similarly, $\mathcal S$ can be calculated at a cost of $\mathcal O({\ndet^2\, \norb}^2)$.
Therefore, asymptotically, calculating $\mathcal H$ is not more expensive than the energy expectation value.

The procedure described so far minimizes the energy with respect to a single ``MO orbital'' of each SD, and different orbitals can be optimized by permuting their order in each determinant.
More generally, we can mix the MO orbitals of each determinant $I$ with a random matrix $U^{(I)} \in \mathrm{SL}(\nele, \mathbb C)$ where $\mathrm{SL}(\nele, \mathbb C)$ is the special linear group, obtaining the transformed orbitals
\begin{equation}
    \phi^{(I)}_{i} \mapsto \sum_{j=1}^{\nele} U_{i,j}^{(I)} \phi^{(I)}_{j},
\end{equation}
and use it to construct an iterative method which eventually optimizes all orbitals, summarized in \cref{alg:algo}.

\begin{algorithm}[H]
\caption{Sketch of the \methodnameacronym algorithm}
\label{alg:algo}
\begin{algorithmic}
\State Initialize Slater determinants
\While{not converged}
    \State Randomly rotate the orbitals $\phi^{(I)} \rightarrow U^{(I)}\phi^{(I)}$
    \State Build $\mathcal H$ and $\mathcal S$
    \State Solve the GEV $\mathcal H v = \epsilon \mathcal S v$
    \State Update $\phi_1^{(I)}$ in each SD $I$
\EndWhile
\end{algorithmic}
\end{algorithm}

We discuss an efficient generalization for the case when the wavefunction \cref{eq:sum_slater_spin} is an Eigenstate of $\hat S_z$ in \externalcref{sup-sec:spin} of the Supplementary Information.\\

\paragraph*{Efficient calculation of the effective matrices.}
We briefly discuss how to compute the matrices $\mathcal H$ and $\mathcal S$ appearing in the EIDOS loop in an efficient and numerically stable way. For notational simplicity, in this section, we drop the determinant index $I$, and we denote by $\ket{\Phi}\coloneq \ket{\Phi^{(I)}},\ket{\Psi}\coloneq\ket{\Phi^{(J)}}$ two generic Slater determinants, and we accordingly define $\phi_j \coloneq \phi_{j}^{(I)}$ and $\psi_j \coloneq \phi_{j}^{(J)}$.
As it is well known, we can write a generalized Wick's theorem for the overlap of two SDs~\cite{Lwdin1955}
\begin{equation}
    \label{eq:det_overlap}
    \braket{\Phi}{\Psi} = \det[A],
\end{equation}
where $A_{ij} = \phi_i^\dagger \psi_j  = \sum_{\mu} \phi^\star_{i,\mu} \psi_{j,\mu}$ is the matrix of pairwise overlaps of the ``MO orbitals''.\\

We consider a generic normal ordered $\nbody$-body operator
\begin{equation}
\label{eq:W_operator}
 \hat W = \sum_{\substack{\xi_1 \dots \xi_\nbody \\ \zeta_1 \dots \zeta_\nbody}}w_{\xi_1 \dots i_\nbody}^{\zeta_1\dots \zeta_\nbody} {\hat  c_{\zeta_\nbody}^\dagger \dots c_{\zeta_1}^\dagger \hat c_{\xi_1} \dots \hat c_{\xi_\nbody}}\\
\end{equation}
and compute the matrix element\begin{equation}
\label{eq:mel_with_rdm}
\mel{\Phi}{\hat W}{\Psi}
= \sum_{\substack{\xi_1 \dots \xi_\nbody \\ \zeta_1 \dots \zeta_\nbody}} w_{\xi_1\dots \xi_\nbody}^{\zeta_1\dots \zeta_\nbody}
\, \rho_{\xi_1\dots \xi_\nbody}^{\zeta_1\dots \zeta_\nbody},
\end{equation}
where
\begin{equation}
\label{eq:rdm_w_contraction}
    \rho_{\xi_1\dots \xi_\nbody}^{\zeta_1\dots \zeta_\nbody}
    \coloneq\mel{\Phi}{\hat  c_{\zeta_\nbody}^\dagger \dots \hat c_{\zeta_1}^\dagger \hat c_{\xi_1} \dots \hat c_{\xi_\nbody}}{\Psi}
\end{equation}
is the $\nbody$-body reduced transition density matrix.

Analytical expressions for the one, two and $\nbody$-body reduced (transition) density matrices and matrix elements of non-orthogonal Slater determinants are well-documented in the literature~\cite{Lwdin1955,Malmqvist1986,Fukutome1988_reshf,Broer1988_mel,Verbeek1991,Koch1993_nonorthog_detopt_singled,Mazziotti1999,Dijkstra2000_gradients_vbt, Tomita2004,Brouder2005,Thom2009,Utsuno2013,laguna_2020_pra_det_mel,burton_2021_mel_paper1,burton2022generalized,Miller2025_reshf_svd}. Here we derive the versions needed for our purpose.

We start with the simplifying assumption that the overlap  $\braket{\Phi}{\Psi}$ is nonzero and write down a generalized Wick's theorem for the $\nbody$-body reduced transition density matrix,
\begin{equation}
\label{eq:rdm_det2}
    \rho_{\xi_1\dots \xi_\nbody}^{\zeta_1\dots \zeta_\nbody}
    = \det[A] \det \bma{
 B^{\xi_1}_{\zeta_1} & \cdots &  B^{\xi_\nbody}_{\zeta_1} \\
 \vdots & \ddots & \vdots\\
 B^{\xi_1}_{\zeta_\nbody} & \cdots &  B^{\xi_\nbody}_{\zeta_\nbody}
 },
\end{equation}
where $B^{\xi}_{\zeta} = \sum_{i,j} \psi_{\xi,i}^\star\, (A^{-1})_{i j}\,\phi_{\zeta,j}$.
We remark that a more general expression for Hartree-Fock-Bogoliubov wavefunctions has been derived~\cite{Bertsch2012,Chen2023_mel_svd_hfb}, which reduces to \cref{eq:rdm_det2} for the special case of SD.
For completeness we include a derivation of \cref{eq:rdm_det2} in \externalcref{sup-proof} of the Supplementary Information.
\cref{eq:rdm_det2} has an overall computational scaling of $\mathcal O(\nbody^3\, \norb^{2\nbody})$ when each determinant of the ${\nbody{\times}\nbody}$ matrices containing the $B$'s is computed explicitly (see e.g. Ref. \cite{laguna_2020_pra_det_mel,burton2022generalized}).\\

\paragraph*{\it Efficiently-contractable formula.} In this work we consider $\nbody \in \{0,1,2,3\}$, for which it is computationally feasible to expand the determinants of \cref{eq:rdm_det2} into their $\nbody!$ terms, and directly contract them with $w$,
\begin{equation}
\label{eq:contraction_part1}
\frac{\mel{\Phi}{\hat W}{\Psi}}{\braket{\Phi}{\Psi}}
=
\sum_{\sigma \in S_\nbody} \mathrm{sgn}(\sigma)
\sum_{\substack{\xi_1 \dots \xi_\nbody \\ \zeta_1 \dots \zeta_\nbody}} w_{\xi_1\dots \xi_\nbody}^{\zeta_1\dots \zeta_\nbody}
 \prod_{t=1}^{\nbody} B^{\xi_t}_{\zeta_{\sigma[t]}},
\end{equation}
where $\mathcal S_{\nbody}$ is the symmetric group and $\sigma[t]$ is the $t$-th element of the permutation $\sigma$.
Whenever $w_{\xi_1\dots \xi_\nbody}^{\zeta_1\dots \zeta_\nbody}$ admits a compact tensor decomposition, we can expect the application of \cref{eq:contraction_part1} to be advantageous.
To give an example of this fact, we consider a factorizable 3-body operator given by $w_{\xi_1 \xi_2 \xi_3}^{\zeta_1 \zeta_2 \zeta_3} = w_{\xi_1 \xi_2}^{\zeta_1 \zeta_2}\, w_{\xi_3}^{\zeta_3}$
as we encounter in \cref{eq:mel_H}.
In this case, an application of the efficiently-contractable formula, \cref{eq:contraction_part1}, requires only $\mathcal O(\norb^{4})$ operations, whereas using \cref{eq:mel_with_rdm,eq:rdm_w_contraction} needs $\mathcal O(\norb^{6})$ operations.\\

\label{sec:svd_version}

For the zero-overlap case $\braket{\Phi}{\Psi} = \det[A] = 0$, the standard approach taken is to do a singular value decomposition (SVD) of $A$~\cite{King1967,Koch1993_nonorthog_detopt_singled,Thom2009,laguna_2020_pra_det_mel,burton_2021_mel_paper1,burton2022generalized,Mahler2021,Chen2023_mel_svd_hfb,Miller2025_reshf_svd}, either generically or using optimized equations for each value of the rank of $A$. Here we follow the former approach.
Taking $A = U \Lambda V^H$ where $\Lambda$ is diagonal and $U,V$ are unitary, we write down the equation of the matrix element of a generic normal ordered \nbody-body operator, given by
\begin{equation}
\label{eq:contraction_part2b}
\begin{aligned}
&\mel{\Phi}{\hat W}{\Psi}
= \det[U] \det[V^H]
\sum_{\sigma \in S_{\nbody}} \mathrm{sgn}(\sigma)
\\ &\quad
\sum_{\substack{\xi_1 \dots \xi_{\nbody} \\ \zeta_1 \dots \zeta_{\nbody}}}
w_{\xi_1\dots \xi_{\nbody}}^{\zeta_1\dots \zeta_{\nbody}}
\sum_{k_1 \dots k_{\nbody}}
T_{k_1 \dots k_{\nbody}}  \prod_{t=1}^{\nbody}
X_{\xi_t, k_t} Y_{k_t, \zeta_{\sigma[t]}},
\end{aligned}
\end{equation}
where $X_{\xi,k} = \sum_{i} \psi_{i,\xi}^\star V_{i,k}$ and $Y_{k,\zeta} = \sum_j U^\star_{j,k} \phi_{j,\zeta}$.
The tensor $T_{k_1\dots k_\nbody}\in \mathbb R$ is the product of all singular values which are not indexed, defined as $T_{k_1 \dots k_\nbody} = \prod_{i \in \{1,\dots \nele\} \setminus \{ k_1, \dots, k_\nbody\}} \Lambda_i$.
We remark that \cref{eq:contraction_part2b} a special case of the more general expression for Hartree-Fock-Bogoliubov wavefunctions given in Eq. 28 of Ref.~\cite{Chen2023_mel_svd_hfb}.
For completeness we include a derivation of \cref{eq:contraction_part2b} in \externalcref{sup-svd_derivation} of the Supplementary Information.
 This equation is correct regardless of the rank of $A$, and it is numerically stable whenever singular values are numerically close to zero.
 The contraction of \cref{eq:contraction_part2b} has a computational cost larger by a factor of $\mathcal O(\nele)$ compared to \cref{eq:contraction_part1}.\\

\label{sec:hess2}

For the matrices $\mathcal H$ and $\mathcal S$, defined in \cref{eq:mathcalH} and \cref{eq:mathcalS}, we need to calculate matrix elements of the form
$\mel*{\Phi} {\hat c_\mu \hat W \hat c_\nu^\dagger}{\Psi}$ for all $\mu,\nu\in \{1,\dots,\norb\}$
where $\hat W$ is a $\nbody$-body operator (see \cref{eq:W_operator}) and thus $\hat c_\mu \hat W \hat c_\nu^\dagger$ is a ${\nbody{+}1}$-body operator.
In the following we show that calculating $\mel*{\Phi} {\hat c_\mu \hat W \hat c_\nu^\dagger}{\Psi}$ for all $\mu,\nu \in \{1,\dots,\norb\}$ has the same asymptotic cost in $\norb$ as $\mel*{\Phi}{\hat W}{\Psi}$.

Using a form of Wick's theorem for normal ordering, or repeated application of the fermionic anticommutation relations,  $\hat c_\mu \hat W \hat c_\nu^\dagger$ can be brought into normal order, resulting in a sum of ${\nbody{+}1}$, $\nbody$ and ${\nbody{-}1}$-body operators.

For $\nbody \geq 1$, in the nonzero-overlap case $\braket{\Phi}{\Psi}\neq 0$, so that we can use \cref{eq:contraction_part1}, we have that
\begin{equation}
\label{eq:mel_H}
\begin{aligned}
&\frac{\mel*{\Phi} {\hat c_\mu \hat W \hat c_\nu^\dagger}{\Psi}}{\braket{\Phi}{\Psi}} = \\
&-
\sum_{\mathclap{\sigma \in S_{\nbody+1}}} \mathrm{sgn}(\sigma)\,
w_{\xi_1\dots \xi_\nbody}^{\zeta_1\dots \zeta_\nbody}
\prod_{t=1}^{\nbody+1} B^{\xi_t}_{\zeta_{\sigma[t]}}\\
&+
\delta_{\mu,\nu}
\sum_{\mathclap{\sigma \in S_{\nbody}}} \mathrm{sgn}(\sigma)\,
w_{\xi_1\dots \xi_\nbody}^{\zeta_1\dots \zeta_\nbody}
\prod_{t=1}^{\nbody} B^{\xi_t}_{\zeta_{\sigma[t]}}\\
&-
\sum_{\mathclap{\substack{\sigma \in S_{\nbody-1}\\u\in\{1,\dots,\nbody\}}}}\mathrm{sgn}(\sigma)\,
w_{\xi_1\dots \xi_\nbody}^{\zeta_1\dots \zeta_{u-1}\, \mu\, \zeta_{u+1} \dots \zeta_\nbody}
\prod_{t=1}^{\nbody} B^{\xi_t}_{\zeta_{\sigma[t]}}\\
&-
\sum_{\mathclap{\substack{\sigma \in S_{\nbody-1}\\v\in\{1,\dots,\nbody\}}}}\mathrm{sgn}(\sigma)\,
w_{\xi_1 \dots \xi_{v-1}\, \nu\, \xi_{v+1} \dots  \xi_\nbody}^{\zeta_1\dots \zeta_\nbody}
\prod_{t=1}^{\nbody} B^{\xi_t}_{\zeta_{\sigma[t]}}\\
&+
\sum_{\mathclap{\substack{\sigma \in S_{\nbody-1}\\u, v \in\{1,\dots,\nbody\}}}}\mathrm{sgn}(\sigma)\,(-1)^{u+v}\,
w_{\xi_1 \dots \xi_{v\text{-}1}\, \nu\, \xi_{v} \dots \xi_{\nbody\text{-}1}}^{\zeta_1\dots \zeta_{u\text{-}1}\, \mu\, \zeta_{u} \dots \zeta_{\nbody\text{-}1}}
\prod_{t=1}^{\nbody\text{-}1} B^{\xi_t}_{\zeta_{\sigma[t]}}
\end{aligned}
\end{equation}
where
we take $\xi_{\nbody+1}\equiv \mu$ and $\zeta_{\nbody+1} \equiv \nu$ for the first term, $\zeta_u \equiv \nu$ for the third and $\xi_v \equiv \mu$ in the fourth,
and use Einstein summation for the remaining $\xi$ and $\zeta$ indices.
The tensors in the equation can be contracted in such a way that the computational cost is $\mathcal O((\nbody+1)!\, \norb^{2\nbody})$.
We point out that for the special case $\nbody=2$ this scaling has been shown using low-rank perturbations by Ref.~\cite{Ikawa1993}.

It can be shown that also in the zero-overlap case, using \cref{eq:contraction_part2b}, the tensors can be contracted in ${\mathcal O((\nbody+1)!\, \norb^{2\nbody}\, \nele)}$.
We provide the full equation in \externalcref{sup-svd_contractions} of the Supplementary Information.

For the special case $\nbody=0$, for $\mathcal S$, we get
\begin{equation}
\label{eq:mel_S}
    \frac{\mel*{\Phi} {\hat c_\mu \hat c_\nu^\dagger}{\Psi}}{\braket*{\Phi}{\Psi}} = -B^{\mu}_{\nu} + \delta_{\mu,\nu} ,
\end{equation} which can be computed in $\mathcal O(\norb^{2})$.

We use \cref{eq:mel_H} and \cref{eq:mel_S} to compute
the matrices \cref{eq:mathcalH} and \cref{eq:mathcalS}.
Given a generic 2-body Hamiltonian $\hat H$, this results in a computational cost of $\mathcal O(\ndet^2 \norb^{4})$, whenever the overlap is nonzero, and $\mathcal O(\ndet^2 \norb^{4}\,\nele)$ otherwise.

\subsection*{Acknowledgements}
We acknowledge insightful discussions with F. Vicentini and L. Viteritti. We thank H. Koch and G. Scuseria for valuable remarks.
This work was supported by the Swiss National Science Foundation under Grant No. 200021\_200336. This research was also supported by SEFRI through Grant No. MB22.00051 (NEQS - Neural Quantum Simulation).

\bibliography{main}

\clearpage
\onecolumngrid

\renewcommand*{\thesection}{\Alph{section}}   
\numberwithin{figure}{section}
\numberwithin{equation}{section}
\renewcommand{\theequation}{{\thesection}\arabic{equation}}
\renewcommand{\thefigure}{{\thesection}\arabic{figure}}

\begin{center}
\textbf{\large Supplementary Information}\\
\end{center}

\section{Implementation details}
\label{sec:impl_details}

We work with the second quantized formulation of the electronic/molecular Hamiltonian in a finite basis set of size $\norb$, given by
\begin{equation}
\label{eq:hamiltonian}
    \hat H
    = \sum_{\substack{\mu\nu\\\sigma}} h_{\mu\nu}\, \hat c_{i\sigma}^\dagger \hat c_{j\sigma}
    + \frac{1}{2} \sum_{\substack{\mu\nu\xi\zeta\\\sigma\sigma^\prime}} h_{\mu\nu\xi\zeta}\, \hat c_{\mu\sigma}^\dagger \hat c_{\xi\sigma^\prime}^\dagger \hat c_{\zeta\sigma^\prime} \hat c_{\nu\sigma}
    + E_{\textrm{nuc}},
\end{equation}
where $\mu,\nu,\xi,\zeta \in \{1,\dots, \norb\}$ and $\sigma,\sigma^\prime \in \{\uparrow, \downarrow\}$.

We get the values for the one- and two-electron integrals, $h_{\mu\nu\xi\zeta}, h_{\mu\nu}$, and $E_{nuc}$,   from PySCF~\cite{Sun2020_pyscf3}. We choose the molecular orbital basis obtained from a Hartree-Fock calculation as an orthonormal basis, and use complex-valued Slater determinants in double precision unless otherwise specified.
We remark that the $\mathcal O(\norb^5)$ scaling of transforming the two-electron integral explicitly into an orthogonal basis could be straightforwardly avoided by including it in \externalcref{main-eq:contraction_part1}/\externalcref{main-eq:contraction_part2b} in the main text and contracting the tensors in $\mathcal O(\norb^4)$.

To improve the overall numerical stability of \methodnameacronym, we normalize the determinants $\ket*{\Psi^{(I)}}$ for computing $\mathcal H$ and $\mathcal S$.
We remark that this can be done without loss of generality, as the linear variational problem of optimizing the coefficients $c_I$ in $\ket {\bar \Psi} = \sum_I c_I \frac{\ket*{\Psi^{(I)}}}{\norm{\ket*{\Psi^{(I)}}}}$, known as NOCI in the literature, is automatically included in the eigenvalue problem \externalcref{main-eq:generalized_eigval} in the main text, and therefore rescaling the determinants at each step do not change the solution of this linear eigenvalue problem.
Furthermore, for additional numerical stability, we orthogonalize the orbitals of every determinant $\ket*{\Psi^{(I)}}$ separately at every step, using a QR decomposition of
the non-square matrices $\bma{\phi^{(I)}_{1}& \phi^{(I)}_{2} & \dots & \phi^{(I)}_{\nele}} \in \mathbb C^{\norb \times \nele}$, leaving the state invariant. This is also a transformation that does not affect the solution of the linear variational problem at each \methodnameacronym step.

\section{Additional Results}
\label{additional_results}

\begin{figure}[h]
\includegraphics[width=0.475\textwidth]{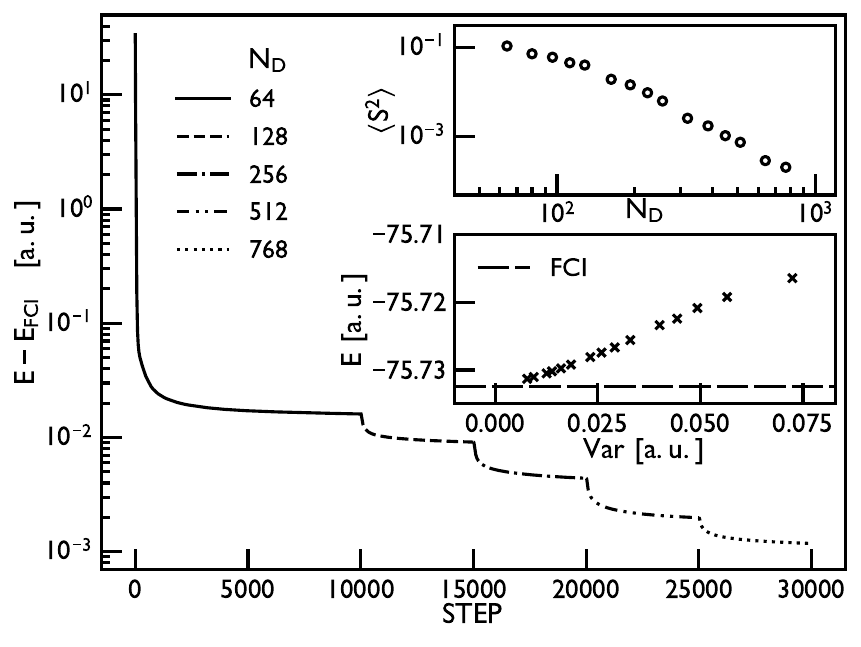}
\caption{\textit{Convergence of the method as a function of the number of iteration steps for the C\textsubscript{2} molecule at equilibrium in the cc-pVDZ basis set}. We compare to the exact FCI energy at the geometry provided in the supplementary material of Ref.~\cite{gao2024distributedfci}. We optimized the first 64 determinants from scratch, and subsequently pre-initialized a part of the determinants using the ones from the previous run upon increasing the number of determinants.
One step comprises the optimization of a single orbital of each of the $\ndet$ determinants.
}
\label{fig:convergence}
\end{figure}

\label{sec:additioal_results}
\paragraph*{Convergence}

In \cref{fig:convergence} we show the convergence to the ground state of the C\textsubscript{2} molecule as a function of the iteration steps, where one step comprises the optimization of one orbital of each determinant.
While in principle we could optimize all Slater determinants from scratch, to reduce the computational cost, we re-use the already optimized Slater Determinants from the previous simulation, adding additional randomly initialized determinants.
We observe a very smooth convergence, with energies that are strictly decreasing at every step. This is expected, as, by definition the energy cannot decrease in our method (unless caused by numerical instabilities).
We further remark that, when using complex-valued determinants we do not observe any orthogonal determinants during the energy optimization, meaning that we can use \externalcref{main-eq:contraction_part1} in the main text and do not have to resort to \externalcref{main-eq:contraction_part2b} in the main text.
However, during testing with real-valued determinants we did observe numerical instabilities requiring the use of the stable svd formula \externalcref{main-eq:contraction_part2b} in the main text.
In the first inset of \cref{fig:convergence} we plot the expectation value of the total spin squared operator, defined as (see  e.g \cite[Chapter 2]{Helgaker2014})
\begin{equation}
\label{eq:s2_def}
    \hat S^2 =
    \frac{1}{2} (\nele^{\downarrow} + \nele^{\uparrow})
    + \frac{1}{4} (\nele^{\downarrow} -\nele^{\uparrow})^2
    - \sum_{t,u=1}^{\norb}
    \hat c^\dagger_{i\uparrow} \hat c_{j\uparrow}
    \hat c^\dagger_{j\downarrow} \hat c_{i\downarrow}
\end{equation}
where
$\ev*{\hat S_z} = \frac{1}{2}(\nele^\uparrow - \nele^\downarrow)$ is fixed in the wavefunction, and $\ev*{\hat S^2} = S(S+1)$ for an Eigenstate of the Hamiltonian.
Using the formulation from \cref{eq:s2_def} we can estimate $\ev*{\hat S^2}$ using a number of operations which scales as $\mathcal O({\norb}^2)$ in $\norb$, the same cost as a generic 1-body operator (also see Sec. 6 of \cite{amos1961single} and supplementary material of Ref. \cite{Miller2025_reshf_svd}).
While for the smaller sizes we observe a small amount of spin contamination for this molecule, the expected value of $S$ goes to 0 using a power-law-like behavior as the number of determinants is increased. Alternatively we could add a penalty term as described in the main text in the context of excited states.
We further compute the energy variance, given by
\begin{equation}
    \textrm{Var} = \expval*{\hat H^2} - \expval*{\hat H}^2.
\end{equation}
Using the efficiently contractable formulae \externalcref{main-eq:contraction_part1,main-eq:contraction_part2b} in the main text (requiring $K=4$) we can compute the expectation value of $\hat H^2$ at an asymptotic cost of $\mathcal O(\ndet^2 {\norb}^4 \nele)$, not more expensive in terms of $\norb$ than the expectation value of the energy itself.
In the second inset of \cref{fig:convergence} we provide an energy/variance plot, of the converged energies from $64$ to $768$ determinants, where we are in the approximately linear regime, suggesting that our states are close to the ground state.

\begin{figure}[h]
\includegraphics[width=0.475\textwidth]{
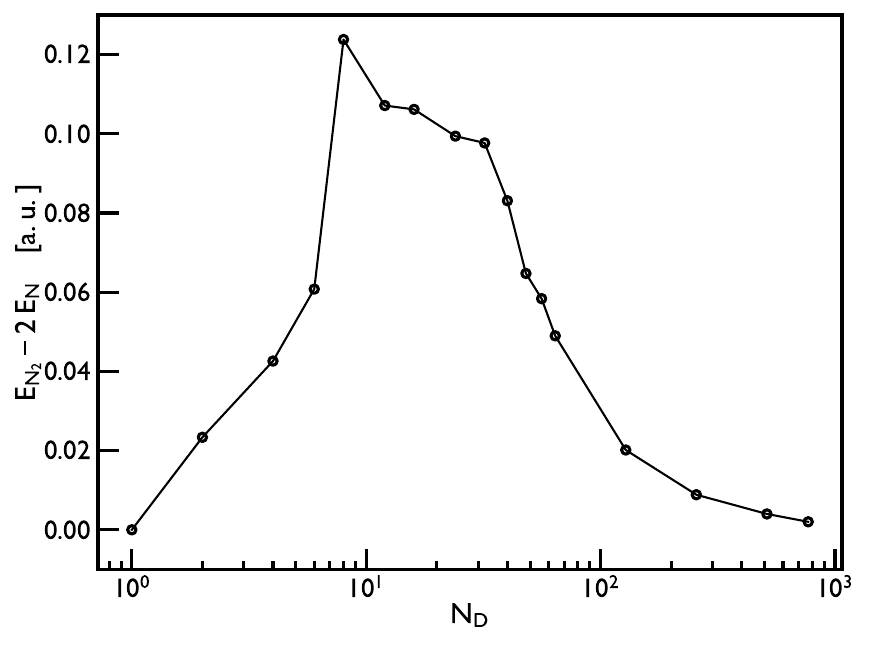
}
\caption{\textit{Size consistency.} We plot the energy difference between the ground state of the N\textsubscript{2} molecule at an approximately infinite bond length compared to twice the energy of a single atom of Nitrogen, using an equal number of determinants.}
\label{fig:size_consistency}
\end{figure}

\paragraph*{Size Consistency}
We plot in \cref{fig:size_consistency} the difference in energy of the ground state of a molecule composed of two N atoms put at very large distance and the sum of the ground state energies of the single atoms, at a fixed number of determinants. The fact that the difference is non zero is a proof of the non-size consistency of the ansatz at fixed number of determinants.

\label{sizeconsistency}

\section{Generalization for unrestricted Slater Determinants}
\label{sec:spin}
 \externalcref{main-alg:algo} in the main text can be generalized to efficiently optimize the wavefunction defined in \externalcref{main-eq:sum_slater_spin,main-eq:slater_det_def} in the main text by modifying it as follows:
At every step we choose to either optimize either the $(I,\uparrow)$ or the $(I,\downarrow)$ part, by sampling $\sigma_I \sim \mathrm{Unif}(\uparrow, \downarrow)$ iid.
Then the vector $v_{I\mu} = \phi^{(I,\sigma_I)}_{1,\mu}$ contains the $\uparrow/\downarrow$ orbitals as chosen, and $\mathcal H$ can be computed as
\begin{equation}
\label{eq:mathcalHspin}
\mathcal H_{I\mu,J\nu} =
\sum_{\mu^\prime \nu^\prime}
\phi^{(I,\bar \sigma_I)\star}_{1,\mu^\prime}
\phi^{(J,\bar \sigma_J)}_{1,\nu^\prime}
\,\bra*{0}
\,\Phi^{(I,\downarrow)\dagger}_{\text{1+}}
{\Phi^{(I,\uparrow)\dagger}_{\text{1+}}
\,
\hat c_{\mu^\prime,\bar \sigma_I}
\hat c_{\mu,\sigma_I}
\,\hat H \,
\hat c_{\nu,\sigma_J}^\dagger
\hat c_{\nu^\prime,\bar \sigma_J}^\dagger
\,
\Phi^{(J\uparrow)}_{\text{1+}}\,\Phi^{(J\downarrow)}_{\text{1+}}}\ket*{0}
\end{equation}
where we use the notation $\bar \uparrow = \downarrow$ and $\bar \downarrow = \uparrow$ for flipping the spin variables. $\mathcal S_{I\mu,J\nu}$ can be calculated analogously, replacing $\hat H$ with the identity.
For the terms in the Hamiltonian which act only on the spin-up or spin-down electrons this factorizes, and we can apply \externalcref{main-eq:mel_H,main-eq:mel_S} in the main text separately to the spin-up and spin-down terms.
This is the case for all the terms in the Hamiltonian \cref{eq:hamiltonian}, with the exception of the 2-body interaction term for $\sigma\neq \sigma^\prime$. For the latter the fermionic operators can be reordered as a product of two hopping terms for spin-up and spin-down, allowing the application of two instances of \externalcref{main-eq:contraction_part1}/\externaleqref{main-eq:contraction_part2b}  in the main text with $\nbody=1$. They can be contracted concurrently, resulting in an overall scaling in $\norb$ of $\mathcal O(\ndet^2{\norb}^4)$/$ O(\ndet^2 {\norb}^4 \nele)$.

\section{Proof of the \nbody-body normal order matrix element formula}
\label{proof}
We recall the well-known Schur Complement formula for the determinant of a block matrix, given by
\begin{equation}
    \label{eq:schur_block_det}
        \det\bma{A & B \\ C & D} = \lim_{\varepsilon \to 0}\det[A + \varepsilon I] \det[D - C (A+\varepsilon I)^{-1} B]
\end{equation}
where we added a infinitesimal diagonal shift $\varepsilon$ so that it can be applied when $A$ is singular.

In the following we prove
\begin{equation}
\label{eq:interaction_generic_normal}
     \mel{\Phi}{\hat c_{\nu_\nbody}^\dagger \dots c_{\nu_1}^\dagger \hat c_{\mu_1} \dots \hat c_{\mu_\nbody} }{\Psi} =
    (-1)^\nbody \det \bma{
    A & \phi_{\nu_1}^T & \dots & \phi_{\nu_\nbody}^T\\
    \psi_{\mu_1}^\star & 0 & \dots & 0 \\
    \vdots & \vdots & \ddots & \vdots\\
    \psi_{\mu_{\nbody}}^\star & 0 & \dots & 0\\
    }
\end{equation}
using
\begin{equation}
\label{eq:interaction_generic_antinormal}
     \mel{\Phi}{\hat c_{\mu_1} \dots \hat c_{\mu_\nbody} \hat c_{\nu_\nbody}^\dagger \dots c_{\nu_1}^\dagger }{\Psi} =
    \det \bma{
    A & \phi_{\nu_1}^T & \dots & \phi_{\nu_\nbody}^T\\
    \psi_{\mu_1}^\star & \delta_{\mu_1 \nu_1} & \dots & \delta_{\mu_1 \nu_\nbody} \\
    \vdots & \vdots & \ddots & \vdots\\
    \psi_{\mu_\nbody}^\star & \delta_{\mu_\nbody \nu_1} & \dots & \delta_{\mu_\nbody \nu_\nbody}\\
    }
\end{equation}
where $\mu_u,\nu_u \in \{1\dots \norb\}$. \cref{eq:interaction_generic_antinormal} follows directly from \cref{main-eq:det_overlap} taking the overlap between the {$(\nele{+}\nbody)$-electron} determinants $c_{\nu_\nbody}^\dagger \dots c_{\nu_1}^\dagger \ket{\Psi}$ and $\hat c_{\mu_\nbody} \dots \hat c_{\mu_1} \ket{\Phi}$.

{\bf Proof}
By induction.

\label{proof_normal_order}
For the trivial $\nbody=0$ case we have that $\braket{\Phi}{\Psi} = \det[A]$.
For $\nbody=1$ we have that
\begin{equation}
\begin{aligned}
     \mel{\Phi}{\hat c_{\nu_1}^\dagger\hat c_{\mu_1}}{\Psi}
     &= \delta_{\mu_1 \nu_1} \braket{\Phi}{\Psi} - \mel{\Phi}{\hat c_{\mu_1} \hat c_{\nu_1}^\dagger}{\Psi} \\
     &= \delta_{\mu_1 \nu_1} \det[A] - \det\bma{A & \phi_{\nu_1}^T\\ \psi_{\mu_1}^\star & \delta_{\mu_1 \nu_1}} \\
     &\overset{(\ref{eq:schur_block_det})}{=} \delta_{\mu_1 \nu_1} \det[A] - \lim_{\varepsilon\to 0} \det[A+\varepsilon I] \big(\delta_{\mu_1 \nu_1} - \psi_{\mu_1}^\star (A+\varepsilon I)^{-1} \phi_{\nu_1}^T\big)\\
     &\overset{(\ref{eq:schur_block_det})}{=} -\det\bma{A & \phi_{\nu_1}^T\\ \psi_{\mu_1}^\star & 0}
\end{aligned}
\end{equation}
The induction assumption is given by :
\begin{equation}
\label{eq:induction_assumption}
\begin{aligned}
     \mel{\Phi}{\hat c_{\nu_\nbody}^\dagger \dots c_{\nu_1}^\dagger \hat c_{\mu_1} \dots \hat c_{\mu_\nbody}}{\Psi}
&\overset{(\ref{eq:interaction_generic_normal})}{=} (-1)^\nbody \det \bma{
    A & \phi_{\nu_1}^T  & \dots & \phi_{\nu_\nbody}^T\\
    \psi_{\mu_1}^\star & 0 & \dots & 0 \\
    \vdots & \vdots & \ddots & \vdots\\
    \psi_{\mu_\nbody}^\star & 0 & \dots & 0\\
    } \\
&\overset{(\ref{eq:schur_block_det})}{=} \lim_{\varepsilon \to 0} \det[A + \varepsilon I] \det[ \bma{\psi_{\mu_t}^\star & \dots & \psi_{\mu_n}^\star}^T (A + \varepsilon I)^{-1} \bma{\phi_{\nu_1}^T & \dots & \phi_{\nu_\nbody}^T}]\\
&= \lim_{\varepsilon \to 0} \det[A + \varepsilon I] \sum_{\sigma \in S_\nbody} \mathrm{sgn}(\sigma) \prod_{t=1}^\nbody (\psi_{\mu_t}^\star (A + \varepsilon I)^{-1} \phi_{\nu_{\sigma[t]}}^T)\\
\end{aligned}
\end{equation}
Next we show the induction step, proving \cref{eq:interaction_generic_normal} for $\nbody>1$, assuming we have already shown it to up to $\nbody-1$.
We use Wick's theorem  to express the antinormal string in normal order
\begin{equation}
\label{eq:a1}
\begin{aligned}
&\mel{\Phi}{\hat c_{\mu_1} \dots \hat c_{\mu_\nbody} \hat c_{\nu_\nbody}^\dagger \dots c_{\nu_1}^\dagger }{\Psi} \\
&\quad = \sum_{P=0}^{\nbody} \sum_{(k_1,l_1) \neq\dots\neq (k_P l_P)} \mel{\Phi}{\normalordL \wick{\hat  c_{\mu_1} \c2 \dots \c1 {\hat c_{\mu_{k_i}}} \c3 \dots c_{\mu_\nbody} \hat c_{\nu_\nbody}^\dagger \c2\dots \c1 {\hat c_{\nu_{l_i}}^\dagger} \c3\dots \hat c_{\nu_1}^\dagger} \normalordR}{\Psi} \\
&\quad =\sum_{P=0}^{\nbody} (-1)^{\nbody-P} \sum_{\substack{|\Omega|=P \\ \Omega\subset \{1,\dots,\nbody\}}} \sum_{\substack{|\Theta|=P \\ \Theta\subset \{1,\dots,\nbody\}}} \sum_{\sigma \in S_{P}} \mathrm{sgn}(\sigma) \big(\prod_{t=1}^{P} \delta_{\mu_{\Omega_t} \nu_{\Theta_{\sigma[t]}}} (-1)^{\Omega_t + \Theta_{\sigma[t]}} \big)
\mel{\Phi}{\hat c_{\nu_{\bar \Theta_{\nbody-P}}}^\dagger \dots \hat c_{\nu_{\bar \Theta_1}}^\dagger \hat c_{\mu_{\bar \Omega_1}} \dots c_{\mu_{\bar \Omega_{\nbody-P}}} }{\Psi} \\
&\quad =(-1)^{\nbody}
\mel{\Phi}{\hat c_{\nu_\nbody}^\dagger \dots \hat c_{\nu_1}^\dagger \hat c_{\mu_1} \dots c_{\mu_\nbody} }{\Psi} \\
&\quad\quad +\sum_{P=1}^{\nbody} (-1)^{\nbody-P} \sum_{\substack{|\Omega|=P \\ \Omega\subset \{1,\dots,\nbody\}}} \sum_{\substack{|\Theta|=P \\ \Theta\subset \{1,\dots,\nbody\}}} \sum_{\sigma \in S_{P}} \mathrm{sgn}(\sigma) \big(\prod_{t=1}^{P} \delta_{\mu_{\Omega_t} \nu_{\Theta_{\sigma[t]}}} (-1)^{\Omega_t + \Theta_{\sigma[t]}} \big)
\mel{\Phi}{\hat c_{\nu_{\bar \Theta_{\nbody-P}}}^\dagger \dots \hat c_{\nu_{\bar \Theta_1}}^\dagger \hat c_{\mu_{\bar \Omega_1}} \dots c_{\mu_{\bar \Omega_{\nbody-P}}} }{\Psi}
\end{aligned}
\end{equation}
where $\wick {\c1 \cdot \quad \c1 \cdot}$ denotes the contraction, $\normalordL\dots\normalordR$ is the notation for normal ordering
and $\bar \Theta, \bar \Omega$ are the complements of the sets $\Theta$ and $\Omega$.
Here we used that the only nonzero terms are those contracting $\hat c$ with $\hat c^\dagger$, thus it is sufficient that we consider all the possible combinations of their subsets of size $N$.
The additional sign factor $(-1)^{\Omega_t + \Theta_{\sigma[t]}}$, comes from swapping the contracted operators to the center so that they are adjacent (after undoing the permutation $\sigma$ which results in the sign $\mathrm{sgn}(\sigma)$).\\
Solving for the normal ordered term $\mel{\Phi}{\hat c_{\nu_\nbody}^\dagger \dots \hat c_{\nu_1}^\dagger \hat c_{\mu_1} \dots c_{\mu_\nbody} }{\Psi}$ it follows that
\begin{equation}
\label{eq:a3}
\begin{aligned}
&\mel{\Phi}{\hat c_{\nu_\nbody}^\dagger \dots \hat c_{\nu_1}^\dagger \hat c_{\mu_1} \dots c_{\mu_\nbody} }{\Psi} \\
&\quad = (-1)^{\nbody} \mel{\Phi}{\hat c_{\mu_1} \dots \hat c_{\mu_\nbody} \hat c_{\nu_\nbody}^\dagger \dots c_{\nu_1}^\dagger }{\Psi} \\
&\quad\quad -\sum_{P=1}^{\nbody} (-1)^{P} \sum_{\substack{|\Omega|=P \\ \Omega\subset \{1,\dots,\nbody\}}} \sum_{\substack{|\Theta|=P \\ \Theta\subset \{1,\dots,P\}}} \sum_{\sigma \in S_{P}} \mathrm{sgn}(\sigma) \big(\prod_{i=1}^{P}\delta_{\mu_{\Omega_t} \nu_{\Theta_{\sigma[t]}}} (-1)^{\Omega_t + \Theta_{\sigma[t]}} \big)
\mel{\Phi}{\hat c_{\nu_{\bar \Theta_{P-\nbody}}}^\dagger \dots \hat c_{\nu_{\bar \Theta_1}}^\dagger \hat c_{\mu_{\bar \Omega_1}} \dots c_{\mu_{\bar \Omega_{P-\nbody}}} }{\Psi}
\end{aligned}
\end{equation}
Using $\prod_{t=1}^{P}(-1)^{\Omega_t + \Theta_{\sigma[t]}}
= (-1)^{\sum_{t=1}^{P} \Omega_t + \sum_{t=1}^{P} \Theta_t} $ to simplify \cref{eq:a3} we have that
\begin{equation}
\label{eq:a4_t2}
\begin{aligned}
&\mel{\Phi}{\hat c_{\nu_P}^\dagger \dots \hat c_{\nu_1}^\dagger \hat c_{\mu_1} \dots c_{\mu_P} }{\Psi} \\
&\quad = (-1)^{P} \mel{\Phi}{\hat c_{\mu_1} \dots \hat c_{\mu_P} \hat c_{\nu_P}^\dagger \dots c_{\nu_1}^\dagger }{\Psi} \\
&\quad\quad -\sum_{P=1}^{\nbody} \sum_{\substack{|\Omega|=P \\ \Omega\subset \{1,\dots,P\}}} \sum_{\substack{|\Theta|=P \\ \Theta\subset \{1,\dots,P\}}}
(-1)^{\sum_{t=1}^{P} \Omega_t + \sum_{t=1}^{P} \Theta_t}
\underbrace{(-1)^{P}\sum_{\sigma \in S_{P}} \mathrm{sgn}(\sigma) \big(\prod_{t=1}^{P}\delta_{\mu_{\Omega_t} \nu_{\Theta_{\sigma[t]}}}\big)}_{=\det\bma{ {-}\delta_{\mu_{\Omega_t} \nu_{\Theta_u}}}_{t,u=1}^P}
\underbrace{\mel{\Phi}{\hat c_{\nu_{\bar \Theta_{P-\nbody}}}^\dagger \dots \hat c_{\nu_{\bar \Theta_1}}^\dagger \hat c_{\mu_{\bar \Omega_1}} \dots c_{\mu_{\bar \Omega_{\nbody-P}}} }{\Psi}}_{\overset{(\ref{eq:induction_assumption})}{=} \substack{\lim_{\varepsilon \to 0} \det[A + \varepsilon I] \hfill \\ \cdot\det\bma{ \psi_{\mu_{\bar \Omega_t}}^\star (A + \varepsilon I)^{-1} \phi_{\nu_{\bar \Theta_u}}^T}_{t,u=1}^{P-\nbody}}} \\
\end{aligned}
\end{equation}
Here we used the induction assumption \cref{eq:induction_assumption} for up to $\nbody-P \leq \nbody-1$.
Next we add and subtract the $P=0$ term of the sum and simplify:
\begin{equation}
\label{eq:a5_t2}
\begin{aligned}
&\mel{\Phi}{\hat c_{\nu_P}^\dagger \dots \hat c_{\nu_1}^\dagger \hat c_{\mu_1} \dots c_{\mu_P} }{\Psi} \\
&\quad = (-1)^{P} \mel{\Phi}{\hat c_{\mu_1} \dots \hat c_{\mu_P} \hat c_{\nu_P}^\dagger \dots c_{\nu_1}^\dagger }{\Psi} \\
&\quad\quad - \underbrace{\lim_{\varepsilon \to 0} \det[A + \varepsilon I] \underbrace{\sum_{P=0}^{\nbody} \sum_{\substack{|\Omega|=P \\ \Omega\subset \{1,\dots,P\}}} \sum_{\substack{|\Theta|=P \\ \Theta\subset \{1,\dots,P\}}}
(-1)^{\sum_{t=1}^{P} \Omega_t + \sum_{t=1}^{P} \Theta_t}
\det\bma{{-} \delta_{\mu_{\Omega_t} \nu_{\Theta_j}}}_{t,u=1}^P
 \det\bma{\psi_{\mu_{\bar \Omega_t}}^\star (A + \varepsilon I)^{-1} \phi_{\nu_{\bar \Theta_j}}^T}_{t,u=1}^{P}}_{=\det\bma{ \psi_{\mu_{\bar \Omega_t}}^\star (A + \varepsilon I)^{-1} \phi_{\nu_{\bar \Theta_j}}^T- \delta_{\mu_i \nu_j}}_{t,u=1}^{P} }}_{\overset{(\ref{eq:interaction_generic_antinormal})\& (\ref{eq:schur_block_det})}{=} (-1)^P \mel{\Phi}{\hat c_{\mu_1} \dots c_{\mu_P} \hat c_{\nu_P}^\dagger \dots \hat c_{\nu_1}^\dagger }{\Psi}} \\
&\quad\quad + \lim_{\varepsilon \to 0} \det[A + \varepsilon I] \det\bma{-\psi_{\mu_t}^\star (A + \varepsilon I)^{-1} \phi_{\nu_u}^T}_{t,u=1}^{P} \\
&=
\lim_{\varepsilon \to 0} \det[A + \varepsilon I] \det\bma{-\psi_{\mu_t}^\star (A + \varepsilon I)^{-1} \phi_{\nu_u}^T}_{t,u=1}^{P}
\overset{(\ref{eq:schur_block_det})}{=} (-1)^P \det \bma{
    A & \psi_{\nu_1}^T & \dots & \psi_{\nu_P}^T \\
    \phi_{\mu_1}^\star & 0 & \dots & 0 \\
    \vdots & \vdots & \ddots & \vdots \\
    \phi_{\mu_n}^\star & 0 & \dots & 0
    }
\end{aligned}
\end{equation}
where in the second step we used Eq.1 from Ref. \cite{marcus1990determinants} for the determinant of a sum two of matrices.

\section{Zero-overlap Formula}
\label{svd_derivation}
Starting from \cref{eq:interaction_generic_normal} we apply \cref{eq:schur_block_det} and singular-value decompose $A = U \mathrm{S} V^H$ where $S$ is diagonal.
Inserting the SVD for both $A$ and $A^{{-}1}$ we have that
\begin{equation}
\label{eq:contraction_part2}
\begin{aligned}
&\mel{\Phi}{\hat W}{\Psi}= \lim_{\varepsilon \to 0} \det[U] \det[S+\varepsilon I] \det[V^H]
\sum_{\substack{\mu_1 \dots \mu_{\nbody} \\ \nu_1 \dots b_{\nbody}}}^{{n_o}} w_{\mu_1\dots \mu_{\nbody}}^{\nu_1\dots \nu_{\nbody}}
\sum_{\substack{
i_1 \dots i_{\nbody} \\
j_1 \dots j_{\nbody} \\
k_1 \dots k_{\nbody}}
}^{n_e}
\sum_{\sigma \in S_\nbody} \mathrm{sgn}(\sigma) \prod_{t=1}^{\nbody}
\psi_{\mu_t,i_t}^\star V_{i_t k_t} (S_{k_t} + \varepsilon)^{-1} U^\star_{j_t k_t} \phi_{\nu_{\sigma[t]} j_t}\\
\end{aligned}
\end{equation}

Here $(S+\varepsilon I)^{-1}$ is applied to an antisymmetric tensor, which is zero if the $\gamma_i$ are not distinct.
This can be shown using the definition in terms of permutations of the generalized Kronecker delta,
\begin{equation}
    \label{eq:antisymm_def}
    \delta_{\nu_1 \dots i_{\nbody}}^{\nu_1 \dots \nu_{\nbody}} = \sum_{\sigma \in S_{\nbody}} \mathrm{sgn}(\sigma)
    \delta_{\nu_1}^{\nu_{\sigma[1]}} \dots \delta_{\nu_{\nbody}}^{\nu_{\sigma[N]}}
\end{equation}
where $S_{\nbody}$ is the symmetric group,
as well as the following identity for moving the antisymmetrization
\begin{equation}
    \label{eq:order_swap}
    \sum_{x_1 \dots x_{\nbody}} \delta_{y_1 \dots y_{\nbody}}^{x_1 \dots x_{\nbody}} A_{x_1}^{z_1} A_{x_2}^{z_2} \dots A_{x_{\nbody}}^{z_\nbody} = \sum_{x^\prime_1 \dots x^\prime_{\nbody}} \delta_{x^\prime_1 \dots x^\prime_{\nbody}}^{z_1 \dots z_{\nbody}} A_{y_1}^{x^\prime_1} A_{y_2}^{x^\prime_2} \dots A_{z_{\nbody}}^{x^\prime_\nbody}
\end{equation}
where
$x_u,y_u \in \{1,2,\dots,Q\}$
and
$z_u,x_u^\prime \in \{1,2,\dots,Q^\prime\}$
for all
$x\in \{1,\dots,Q\}$
and arbitrary $Q,Q^\prime$.
Then we can apply the following identity, which allows us to remove the singularity when one of the singular values is zero, taking the limit $\varepsilon \to 0$:
\begin{equation}
\label{eq:s_prod_identity}
\begin{aligned}
&    \lim_{\epsilon \to 0} \Big(\prod_{k=1}^{\nele} (\Lambda_k+\epsilon) \Big)\,
    \delta_{i_1\dots i_{\nbody}}^{j_1 \dots j_{\nbody}} \,
    \prod_{u=1}^\nbody
    (\Lambda_{i_u}+\epsilon)^{-1}
=
\delta_{i_1\dots i_{\nbody}}^{j_1 \dots j_{\nbody}} \,
\underbrace{
\prod_{\ell \in \{1,\dots \nele\} \setminus \{k_1, \dots, k_{\nbody}\}} \Lambda_\ell
}_{
\eqqcolon
T_{k_1, k_2,\dots, k_{\nbody}}
}
\end{aligned}
\end{equation}
where the generalized Kronecker delta is a placeholder for an arbitrary antisymmetric tensor, and we implicitly define the tensor $T_{k_1, k_2,\dots, k_{\nbody}} \in \mathbb R$ as the product of all singular values which are not indexed.

Finally, applying \cref{eq:s_prod_identity} to \cref{eq:contraction_part2}
 we get
\begin{equation}
\begin{aligned}
&\mel{\Phi}{\hat W}{\Psi}
&= \det[U] \det[V^H]
\sum_{\substack{\mu_1 \dots \mu_{\nbody} \\ \nu_1 \dots \nu_{\nbody}}}^{{n_o}} w_{\mu_1\dots \mu_{\nbody}}^{\nu_1\dots \nu_{\nbody}}
\sum_{\substack{i_1 \dots i_{\nbody} \\ j_1 \dots j_{\nbody} \\ k_1 \dots k_{\nbody}}}^{n_e}
T_{k_1 \dots k_{\nbody}} \sum_{\sigma \in S_{\nbody}} \mathrm{sgn}(\sigma) \prod_{t=1}^{P}
\psi_{i_t,\mu_t}^\star
V_{i_t k_t}
U^\star_{k_t j_t}
\phi_{j_t, \nu_{\sigma[t]}}
\end{aligned}
\end{equation}
 We remark that using \cref{eq:order_swap} we can move the antisymmetrization to one of the other variables $\mu$, $\zeta$ or $\gamma$ as is convenient.
\section{Matrix elements (SVD Version)}
\label{svd_contractions}
The analog of \externalcref{main-eq:mel_H}  in the main text for the zero-overlap case, using \externalcref{main-eq:contraction_part2b}  in the main text instead of \externalcref{main-eq:contraction_part1}  in the main text is given by
\begin{equation}
\label{eq:mel_H_svd}
\begin{aligned}
&\frac{\mel*{\Phi} {\hat c_\mu \hat W \hat c_\nu^\dagger}{\Psi}}{\det[U]\det[V^H]} = \\
&-
\sum_{\mathclap{\sigma \in S_{\nbody+1}}} \mathrm{sgn}(\sigma)\,
w_{\xi_1\dots \xi_\nbody}^{\zeta_1\dots \zeta_\nbody}
\sum_{k_1\dots k_{\nbody+1}} T_{k_1\dots k_{\nbody+1}}
\prod_{t=1}^{\nbody+1} \tilde B_{\xi_t,\zeta_{\sigma[t]},k_t}\\
&+
\delta_{\mu,\nu}
\sum_{\mathclap{\sigma \in S_{\nbody}}} \mathrm{sgn}(\sigma)\,
w_{\xi_1\dots \xi_\nbody}^{\zeta_1\dots \zeta_\nbody}
\sum_{k_1\dots k_{\nbody}} T_{k_1\dots k_{\nbody}}
\prod_{t=1}^{\nbody} \tilde B_{\xi_t,\zeta_{\sigma[t]},k_t}\\
&-
\sum_{\mathclap{\substack{\sigma \in S_{\nbody-1}\\u\in\{1,\dots,\nbody\}}}}\mathrm{sgn}(\sigma)\,
w_{\xi_1\dots \xi_\nbody}^{\zeta_1\dots \zeta_{u-1}\, \mu\, \zeta_{u+1} \dots \zeta_\nbody}
\sum_{k_1\dots k_{\nbody}} T_{k_1\dots k_{\nbody}}
\prod_{t=1}^{\nbody} \tilde B_{\xi_t,\zeta_{\sigma[t]},k_t}\\
&-
\sum_{\mathclap{\substack{\sigma \in S_{\nbody-1}\\v\in\{1,\dots,\nbody\}}}}\mathrm{sgn}(\sigma)\,
w_{\xi_1 \dots \xi_{v-1}\, \nu\, \xi_{v+1} \dots\xi_\nbody}^{\zeta_1\dots \zeta_\nbody}
\sum_{k_1\dots k_{\nbody}} T_{k_1\dots k_{\nbody}}
\prod_{t=1}^{\nbody} \tilde B_{\xi_t,\zeta_{\sigma[t]},k_t}\\
&+
\sum_{\mathclap{\substack{\sigma \in S_{\nbody-1}\\u, v \in\{1,\dots,\nbody\}}}}\mathrm{sgn}(\sigma)\,(-1)^{u+v}\,
w_{\xi_1 \dots \xi_{v\text{-}1}\, \nu\, \xi_{v} \dots \xi_{\nbody\text{-}1}}^{\zeta_1\dots \zeta_{u\text{-}1}\, \mu\, \zeta_{u} \dots \zeta_{\nbody\text{-}1}}
\sum_{k_1\dots k_{\nbody-1}} T_{k_1\dots k_{\nbody}}
\prod_{t=1}^{\nbody\text{-}1} \tilde B_{\xi_t,\zeta_{\sigma[t]},k_t}
\end{aligned}
\end{equation}
where $\tilde B_{\xi,\zeta,k} \coloneqq \sum_{i,j} \psi_{\xi,i}^\star\, V^\star_{i,k} U_{j,k}\,\phi_{\zeta,j}$,
we take $\xi_{\nbody+1}\equiv \mu$ and$\zeta_{\nbody+1} \equiv \nu$ for the first term, $\zeta_u \equiv \nu$ for the third and $\xi_v \equiv \mu$ in the fourth,
and use Einstein summation for the remaining $\xi$ and $\zeta$ indices.
The variables $U,V,T$ are defined as in the discussion of \externalcref{main-eq:contraction_part2b}  in the main text.
The tensors in the equation can be contracted in such a way that the computational cost is $\mathcal O((\nbody+1)!\, \norb^{2\nbody} \,\nele)$.

\end{document}